\documentclass[12pt]{article}
\usepackage{amsmath,amssymb,amsthm,amsxtra,overpic,bbm,bm,epsfig,ulem,subfigure}
\usepackage{color}
\usepackage{hyperref}
\usepackage{mathrsfs}
\usepackage{graphicx}
\usepackage{comment}
\usepackage{float}
\usepackage{cite}
\usepackage{url}
\usepackage{slashed,stmaryrd}

\def\thefootnote{\fnsymbol{footnote}}

\begin{document}

\vspace{0.2cm}

\begin{center}
{\large\bf Family-separated seesaw relations of Majorana neutrinos}
\end{center}

\vspace{0.2cm}

\begin{center}
{\bf Zhi-zhong Xing$^{1,2}$}
\footnote{E-mail: xingzz@ihep.ac.cn}
\\
{\small $^{1}$Institute of High Energy Physics and School of Physical Sciences, \\
University of Chinese Academy of Sciences, Beijing 100049, China \\
$^{2}$Center of High Energy Physics, Peking University, Beijing 100871, China}
\end{center}

\vspace{2cm}
\begin{abstract}
Given the canonical seesaw mechanism as a most natural extension
of the standard model in its neutrino sector, we find out a special
but brand new solution to the exact seesaw equation:
$m^{}_i/M^{}_i = - R^2_{\alpha i}/U^2_{\alpha i}$ for the masses and
flavor mixing matrix elements of light and heavy Majorana neutrinos
of the $i$-th family (for $i = 1, 2, 3$ and $\alpha = e, \mu, \tau$).
This family-separated seesaw scenario allows us to establish simple
relations between the original seesaw parameters and the active
degrees of freedom, and thus offers a number of testable predictions
in neutrino phenomenology.
\end{abstract}

\newpage

\def\thefootnote{\arabic{footnote}}
\setcounter{footnote}{0}
\setcounter{equation}{0}
\setcounter{figure}{0}

\section{Motivation}

The standard model (SM) of particle physics is widely recognized as the most
successful quantum field theory in describing the fundamental properties
of electroweak and strong
interactions, but it remains incomplete for several reasons~\cite{Xing:2019vks},
especially because it leaves no room for accommodating tiny neutrino masses
and significant lepton flavor mixing effects that have been discovered in a number
of milestone experiments of neutrino oscillations~\cite{ParticleDataGroup:2024cfk}.
But the unique dimension-five Weinberg operator of the SM effective field
theory can provide a natural and costless interpretation of why the three known
active neutrinos should have the Majorana nature and sufficiently small
masses~\cite{Weinberg:1979sa}, with no need for invoking any details of
underlying new physics.

The simplest and well-motivated way of going beyond the SM to realize the Weinberg
operator is the canonical seesaw mechanism which
introduces three right-handed neutrino fields and their kinetic terms, the
Yukawa interactions of left- and right-handed neutrino fields with the SM Higgs
field, and the self-interaction term of right-handed neutrino fields and their
charge-conjugated counterparts --- a lepton-number-violating Majorana mass
term as follows~\cite{Minkowski:1977sc}:
\begin{eqnarray}
{\cal L}^{}_{\rm seesaw} = \overline{N^{}_{\rm R}} \hspace{0.08cm}
{\rm i} \hspace{0.03cm} \slashed{\partial} N^{}_{\rm R} - \left[
\overline{\ell^{}_{\rm L}} \hspace{0.08cm} Y^{}_\nu \widetilde{H}
N^{}_{\rm R} + \frac{1}{2} \hspace{0.05cm}
\overline{(N^{}_{\rm R})^c} \hspace{0.05cm} M^{}_{\rm R} N^{}_{\rm R}
+ {\rm h.c.} \right] \; ,
\label{1}
%     (1)
\end{eqnarray}
in which $N^{}_{\rm R} = (N^{}_{e \rm R} \hspace{0.24cm}
N^{}_{\mu \rm R} \hspace{0.24cm} N^{}_{\tau \rm R})^T$ denotes
the column vector of three right-handed neutrino fields belonging to
the $\rm SU(2)^{}_{\rm L}$ singlets,
$\ell^{}_{\rm L} = (\nu^{}_{\rm L} \hspace{0.24cm}
l^{}_{\rm L})^T$ is an $\rm SU(2)^{}_{\rm L}$ lepton
doublet with $\nu^{}_{\rm L} = (\nu^{}_{e \rm L} \hspace{0.24cm}
\nu^{}_{\mu \rm L} \hspace{0.24cm} \nu^{}_{\tau \rm L})^T$ and
$l^{}_{\rm L} = (l^{}_{e \rm L} \hspace{0.24cm} l^{}_{\mu \rm L}
\hspace{0.24cm} l^{}_{\tau \rm L})^T$,
$\widetilde{H} \equiv {\rm i} \sigma^{}_2 H^*$
with $H = (\phi^+ \hspace{0.24cm} \phi^0)^T$ being the Higgs doublet
of the SM, $(\nu^{}_{\rm L})^c$ and $(N^{}_{\rm R})^c$ represent the
charge-conjugated counterparts of $\nu^{}_{\rm L}$ and
$N^{}_{\rm R}$, $Y^{}_\nu$ is the Yukawa coupling matrix of massive
neutrinos, and $M^{}_{\rm R}$ is a symmetric Majorana mass matrix.
In this straightforward gauge-invariant extension of the SM, the
Weinberg operator can easily be achieved after those new and heavy
degrees of freedom far above the electroweak scale are integrated
out~\cite{Broncano:2002rw,Antusch:2006vwa,Zhang:2021jdf}.
Here we focus on the seesaw framework itself, and arrive at the
lepton mass terms
\begin{eqnarray}
-{\cal L}^{}_{\rm mass} = \overline{l^{}_{\rm L}} \hspace{0.08cm}
M^{}_l \hspace{0.05cm} l^{}_{\rm R} + \frac{1}{2} \hspace{0.05cm}
\overline{\displaystyle \left[ \nu^{}_{\rm L} \hspace{0.2cm}
(N^{}_{\rm R})^c\right]} \left(\begin{matrix} {\bf 0} & M^{}_{\rm D} \cr
M^T_{\rm D} & M^{}_{\rm R} \end{matrix}\right)
\left[\begin{matrix} (\nu^{}_{\rm L})^c \cr N^{}_{\rm R} \end{matrix}
\right] + {\rm h.c.} \;
\label{2}
%     (2)
\end{eqnarray}
after the Higgs field $\phi^0$ acquires its vacuum expectation value
$\langle \phi^0\rangle \simeq 174~{\rm GeV}$~\cite{Xing:2011zza},
where $M^{}_l \equiv Y^{}_l \langle \phi^0\rangle$ with $Y^{}_l$
being the charged-lepton Yukawa coupling matrix,
$M^{}_{\rm D} \equiv Y^{}_\nu \langle \phi^0\rangle$, and $\bf 0$ denotes
the $3\times 3$ zero matrix. Without loss of any generality, one may
choose $M^{}_l = D^{}_l \equiv {\rm Diag}\{ m^{}_e, m^{}_\mu, m^{}_\tau\}$
for the charged leptons. Then lepton flavor mixing in weak
charged-current interactions arises purely from a unitary
transformation of the flavor basis of six neutrino fields in
Eq.~(\ref{2}) into their mass basis:
\begin{eqnarray}
-{\cal L}^{}_{\rm cc} = \frac{g}{\sqrt{2}} \hspace{0.06cm}
\overline{\big(\begin{matrix} e & \mu & \tau\end{matrix}\big)^{}_{\rm L}}
\hspace{0.08cm} \gamma^\mu \left[ U \left( \begin{matrix} \nu^{}_{1}
\cr \nu^{}_{2} \cr \nu^{}_{3} \cr\end{matrix}
\right)^{}_{\hspace{-0.08cm} \rm L}
+ R \left(\begin{matrix} N^{}_1 \cr N^{}_2 \cr N^{}_3
\cr\end{matrix}\right)^{}_{\hspace{-0.08cm} \rm L} \hspace{0.05cm} \right]
W^-_\mu + {\rm h.c.} \; ,
\label{3}
%     (3)
\end{eqnarray}
where $g$ is the weak coupling constant, $U^{}_{\alpha i}$ denote the
elements of the Pontecorvo-Maki-Nakagawa-Sakata (PMNS) matrix associated
with the light Majorana neutrinos~\cite{Pontecorvo:1957cp,Maki:1962mu},
and $R^{}_{\alpha i}$ represent the elements of a PMNS-like matrix
responsible for active-sterile flavor mixing and relevant to the heavy
Majorana neutrinos (for $\alpha = e, \mu, \tau$ and
$i = 1, 2, 3$)~\cite{Xing:2007zj}. Note that $U$ and $R$ are correlated
with each other via both the exact seesaw relation and a partial unitary
condition
%%%%%%%%%%%%%%%%%%%%%%%%%%%%%%%%%%%%%%%%%%%%%%%%%%%%%%%%%%%%%%%%%%%
\footnote{The reason for these two correlations is that $U$ and $R$
belong respectively to the left- and right-upper sub-matrices of the
$6 \times 6$ unitary matrix used to diagonalize the $6 \times 6$
overall neutrino mass matrix in Eq.~(\ref{2}). So the PMNS matrix $U$ 
itself is not exactly unitary, and its slight deviation from a unitary 
matrix $U^{}_0$ can be expressed as $U = A \cdot U^{}_0$ with $A$ being 
a lower triangular matrix which slightly deviates from the identity 
matrix $\bf 1$~\cite{Xing:2007zj,Xing:2011ur}. The meaning of a 
``partial" unitary condition in Eq.~(\ref{4}) is simply that 
$U^\dagger U + R^\dagger R = {\bf 1}$ does not hold in the seesaw 
framework.}:
%%%%%%%%%%%%%%%%%%%%%%%%%%%%%%%%%%%%%%%%%%%%%%%%%%%%%%%%%%%%%%%%%%%%%
\begin{eqnarray}
U D^{}_\nu U^T + R D^{}_N R^T = {\bf 0} \; , \quad
U U^\dagger + R R^\dagger = {\bf 1} \; ,
\label{4}
%     (4)
\end{eqnarray}
where $D^{}_\nu = {\rm Diag}\{m^{}_1, m^{}_2, m^{}_3\}$ and
$D^{}_N = {\rm Diag}\{M^{}_1, M^{}_2, M^{}_3\}$ with $m^{}_i$ and
$M^{}_i$ being the corresponding masses of $\nu^{}_i$ and $N^{}_i$
(for $i = 1, 2, 3$), and $\bf 1$ denotes the $3\times 3$ identity matrix.
Unfortunately, solving the above seesaw equation in an analytically
exact manner has been a big challenge as $U$ and $R$ are nonlinearly
entangled and involve many free parameters
%%%%%%%%%%%%%%%%%%%%%%%%%%%%%%%%%%%%%%%%%%%%%%%%%%%%%%%%%%%%%%%%%%
\footnote{A remarkable attempt along this line of thought has recently
been made with the help of some safe analytical approximations for
$U$ and $R$~\cite{Xing:2024gmy}, but the relevant results are so
complicated that it remains difficult to find sufficiently
interesting and testable phenomenological applications for them.}.
%%%%%%%%%%%%%%%%%%%%%%%%%%%%%%%%%%%%%%%%%%%%%%%%%%%%%%%%%%%%%%%%%%
In this case whether $m^{}_i$ and $M^{}_i$ (or $U^{}_{\alpha i}$
and $R^{}_{\alpha i}$) in the $i$-th family of massive neutrinos
may possess a simple linear relation has been unclear.

The present work is intended to report a particular but brand new
solution to the exact seesaw equation in Eq.~(\ref{4}):
$m^{}_i/M^{}_i = -R^{2}_{\alpha i}/U^{2}_{\alpha i}$ for the masses
and flavor mixing matrix elements of light and heavy Majorana neutrinos
of the $i$-th family (for $i = 1, 2, 3$ and $\alpha = e, \mu, \tau$).
Such a novel solution leads us to a family-separated seesaw (FSS) scenario,
which makes it possible to calculate the flavor parameters in $D^{}_N$
and $R$ with those light and active degrees of freedom in $D^{}_\nu$ and
$U$ (or vice versa). In this work we are going to outline the salient 
features of the FSS scenario and explore its predictability and testability
in neutrino phenomenology. 

\section{The FSS scenario}

\begin{figure}[t]
\begin{center}
\includegraphics[width=14.5cm]{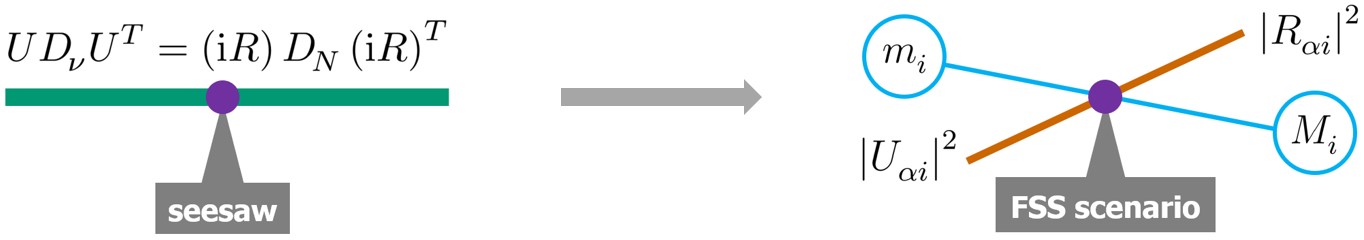}
\caption{A schematic illustration of the family-separated seesaw (FSS)
scenario as a particular but brand new solution to the exact seesaw
equation $U D^{}_{\nu} U^T = \left({\rm i} R\right) D^{}_{N} \left({\rm i}
R\right)^T$ regarding the light and heavy Majorana neutrinos (for
$\alpha = e, \mu, \tau$ and $i = 1, 2, 3$).}
\label{fig1}
\end{center}
\end{figure}

The exact seesaw formula and the unitarity condition shown in Eq.~(\ref{4})
can be explicitly expressed as
\begin{eqnarray}
&& \sum^3_{i=1} \left(m^{}_i U^{}_{\alpha i} U^{}_{\beta i} +
M^{}_i R^{}_{\alpha i} R^{}_{\beta i}\right) = 0 \; ,
\nonumber \\
&&
\sum^3_{i=1} \left(U^{}_{\alpha i} U^{*}_{\beta i} +
R^{}_{\alpha i} R^{*}_{\beta i}\right)
= \delta^{}_{\alpha\beta} \; , \hspace{1.8cm}
\label{5}
%     (5)
\end{eqnarray}
in which the Greek subscripts $\alpha$ and $\beta$ run over the three
lepton flavor indices $e$, $\mu$ and $\tau$. Motivated by the
well-known Occam's Razor (simplicity) principle, we conjecture that the
three families of light and heavy Majorana neutrinos may have their own
seesaw relations. In other words, we expect that there may exist a
particular but instructive solution to the above seesaw equation:
\begin{eqnarray}
m^{}_i U^{}_{\alpha i} U^{}_{\beta i} + M^{}_i R^{}_{\alpha i}
R^{}_{\beta i} = 0 \;
\label{6}
%     (6)
\end{eqnarray}
itself instead of a summation over the family index $i$ (for
$\alpha, \beta = e, \mu, \tau$ and $i = 1, 2, 3$), from which
the brand new FSS relations
\begin{eqnarray}
\frac{R^{}_{e i}}{U^{}_{e i}} =
\frac{R^{}_{\mu i}}{U^{}_{\mu i}} =
\frac{R^{}_{\tau i}}{U^{}_{\tau i}} = {\rm i} \sqrt{\frac{m^{}_i}
{M^{}_i}} \;
\label{7}
%     (7)
\end{eqnarray}
can be easily achieved. It is obvious that the ratios
$R^{}_{\alpha i}/U^{}_{\alpha i}$ are insensitive to any redefinition
of the phases of three charged-lepton fields, but they impose some
strong constraints on the phases of the flavor mixing matrix elements
$R^{}_{\alpha i}$ and $U^{}_{\alpha i}$ as follows:
\begin{eqnarray}
\arg R^{}_{\alpha i} = \arg U^{}_{\alpha i}
+ \frac{4n + 1}{2} \pi \;
\label{8}
%     (8)
\end{eqnarray}
with $n$ being an arbitrary integer. A rephasing-invariant version
of the FSS relations in Eq.~(\ref{7}) is
\begin{eqnarray}
m^{}_i \left|U^{}_{\alpha i}\right|^2 =
M^{}_i \left|R^{}_{\alpha i}\right|^2 \; ,
\label{9}
%     (9)
\end{eqnarray}
corresponding to three separate seesaw relations of the three Majorana
neutrino families in their mass basis. Fig.~\ref{fig1} offers an
illustration of the FSS scenario reduced from the exact seesaw equation
as its simplest special solution. In this case, each of the three
neutrino families has its own seesaw relation either between $m^{}_i$
and $M^{}_i$ or between $U^{}_{\alpha i}$ and $R^{}_{\alpha i}$.
Given the prerequisite $M^{}_i \gg \langle\phi^0\rangle$ to assure the
naturalness of a seesaw mechanism, together with ${\cal O}(0.1) \lesssim
|U^{}_{\alpha i}| < 1$ extracted from current neutrino oscillation
data~\cite{ParticleDataGroup:2024cfk}, we conclude that the
smallness of $m^{}_i$ is definitely attributed to the strongly
suppressed active-sterile flavor mixing effects as described by the
smallness of $\left|R^{}_{\alpha i}\right|$ (i.e., very weak Yukawa
interactions of the neutrino fields because of
$Y^{}_\nu \propto R$~\cite{Xing:2023adc}).
%%%%%%%%%%%%%%%%%%%%%%%%%%%%%%% Figure 1 %%%%%%%%%%%%%%%%%%%%%%%%%%%%%%%
%%%%%%%%%%%%%%%%%%%%%%%%%%%%%%%%%%%%%%%%%%%%%%%%%%%%%%%%%%%%%%%%%%%%%%%%

Note that the FSS relations in Eq.~(\ref{7}) are compatible
with possible $\mu$-$\tau$ reflection symmetries associated with the
light and heavy Majorana neutrinos, 
\begin{eqnarray}
{\cal P} \left(U D^{}_\nu U^T\right) {\cal P}^T = 
\left(U D^{}_\nu U^T\right)^* \; , \quad 
{\cal P} \left(R D^{}_N R^T\right) {\cal P}^T =
\left(R D^{}_N R^T\right)^* \; ,
\label{10}
%     (10)
\end{eqnarray}
where ${\cal P} = {\cal P}^T = {\cal P}^\dagger$ with
${\cal P}^{}_{11} = {\cal P}^{}_{23} = {\cal P}^{}_{32} = 1$,
and the other six elements of ${\cal P}$ are all zero. The most 
interesting consequences of such a minimal flavor symmetry are
$\left|U^{}_{\mu i}\right| = \left|U^{}_{\tau i}\right|$ and 
$\left|R^{}_{\mu i}\right| = \left|R^{}_{\tau i}\right|$ (for $i = 1, 2, 3$).
Given the fact that an approximate $\mu$-$\tau$ reflection symmetry of $U$ is 
actually supported by current neutrino oscillation 
data~\cite{Xing:2015fdg,Xing:2022oob,Xing:2022uax}, it can certainly be applied 
to $R$ to further constrain the active-sterile flavor mixing effects 
in the FSS scenario.

In this new seesaw scenario, the Jarlskog-like rephasing invariants of CP
violation associated with the three active Majorana neutrinos at low
energies can be expressed as
\begin{eqnarray}
{\cal J}^{ij}_{\alpha\beta} \hspace{-0.2cm} & \equiv & \hspace{-0.2cm}
{\rm Im}\left(U^{}_{\alpha i} U^{}_{\beta j} U^*_{\alpha j}
U^*_{\beta i}\right) = \frac{M^{}_i M^{}_j}{m^{}_i m^{}_j}
{\rm Im}\left(R^{}_{\alpha i} R^{}_{\beta j} R^*_{\alpha j}
R^*_{\beta i}\right) \; ,
\nonumber \\
{\cal V}^{ij}_{\alpha\beta} \hspace{-0.2cm} & \equiv & \hspace{-0.2cm}
{\rm Im}\left(U^{}_{\alpha i} U^{}_{\beta i} U^*_{\alpha j}
U^*_{\beta j}\right) = \frac{M^{}_i M^{}_j}{m^{}_i m^{}_j}
{\rm Im}\left(R^{}_{\alpha i} R^{}_{\beta i} R^*_{\alpha j}
R^*_{\beta j}\right) \; , \hspace{0.4cm}
\label{11}
%     (11)
\end{eqnarray}
which describe the CP violating effects in the lepton-number-conserving
and lepton-number-violating processes\cite{Luo:2011mm,Xing:2013ty,
Xing:2013woa,Wang:2021rsi,Xing:2025zqt}, respectively. It is therefore
possible to probe the original seesaw phase parameters or their combinations
hidden in $R$ via the measurements of CP violation in normal neutrino
oscillations and (in principle) neutrino-antineutrino oscillations.

But which of the two kinds of rephasing invariants defined in Eq.~(\ref{11}),
${\cal J}^{ij}_{\alpha\beta}$ or ${\cal V}^{ij}_{\alpha\beta}$, can be 
directly related to CP violation in the decays of heavy Majorana neutrinos 
and thus to leptogenesis~\cite{Fukugita:1986hr}? To answer this question, 
let us simply consider the flavor-independent CP-violating asymmetry
$\varepsilon^{}_i$ between the decay mode ${N}^{}_i \to \ell^{}_\alpha
+ H$ and its CP-conjugated process ${N}^{}_i \to \overline{\ell^{}_\alpha}
+ \overline{H}$ far above the electroweak symmetry breaking scale
$\langle \phi^0\rangle \simeq 174~{\rm GeV}$~\cite{Luty:1992un,Covi:1996wh,
Xing:2023kdj}. We arrive at
\begin{eqnarray}
\varepsilon^{}_i \hspace{-0.2cm} & \equiv & \hspace{-0.2cm}
\sum_\alpha \left[
\frac{\Gamma({N}^{}_i \to \ell^{}_\alpha + H)
- \Gamma({N}^{}_i \to \overline{\ell^{}_\alpha} +
\overline{H})}{\displaystyle \sum_\alpha \left[\Gamma({N}^{}_i \to
\ell^{}_\alpha + H) + \Gamma({N}^{}_i \to \overline{\ell^{}_\alpha}
+ \overline{H})\right]}\right]
\nonumber \\
\hspace{-0.2cm} & = & \hspace{-0.2cm}
\frac{1}{\displaystyle 8\pi \sum_\alpha \left|R^{}_{\alpha i}\right|^2}
\sum_{k \neq i} \left[\frac{M^2_k}{\langle \phi^0\rangle^2}
\hspace{0.05cm} {\rm Im}
\left(\sum_\alpha R^*_{\alpha i} R^{}_{\alpha k}\right)^2
\xi\left(x^{}_{k i}\right)\right]
\nonumber \\
\hspace{-0.2cm} & = & \hspace{-0.2cm}
\frac{-1}{\displaystyle 8\pi \sum_\alpha \left|U^{}_{\alpha i}\right|^2}
\sum_{k \neq i} \left[\frac{m^{}_k M^{}_k}{\langle \phi^0\rangle^2}
\left(\sum_\alpha {\cal V}^{ik}_{\alpha\alpha} + 2
\sum_{\alpha < \beta} {\cal V}^{jk}_{\alpha\beta} \right)
\xi\left(x^{}_{k i}\right)\right] \; , \hspace{0.8cm}
\label{12}
%     (12)
\end{eqnarray}
where $\alpha$ and $\beta$ run over $(e, \mu, \tau)$, and
$\xi\left(x^{}_{ki}\right) = \sqrt{x^{}_{ki}} \left\{1 +
1/\left(1 - x^{}_{ki}\right) + \left(1 + x^{}_{ki}\right)
\ln\left[x^{}_{ki} / \left(1 + x^{}_{ki}\right)\right] \right\}$
denotes the loop function with $x^{}_{ki} \equiv {M}^2_{k}/{M}^2_i$
(for $i, k = 1, 2, 3$ but $k \neq i$). This result shows that the
asymmetry $\varepsilon^{}_i$ is not directly related to
${\cal J}^{ij}_{\alpha\beta}$, and hence it is not directly
correlated with CP violation in normal neutrino oscillations with
lepton number conservation in the FSS scenario. Nevertheless, one
may see that $\varepsilon^{}_i$ is linearly related to a summation
of the terms proportional to $m^{}_k M^{}_k/\langle \phi^0\rangle^2$
--- a perfect reflection of the FSS relations between the light and
heavy Majorana neutrinos.

\begin{figure}[t]
\begin{center}
\includegraphics[width=9cm]{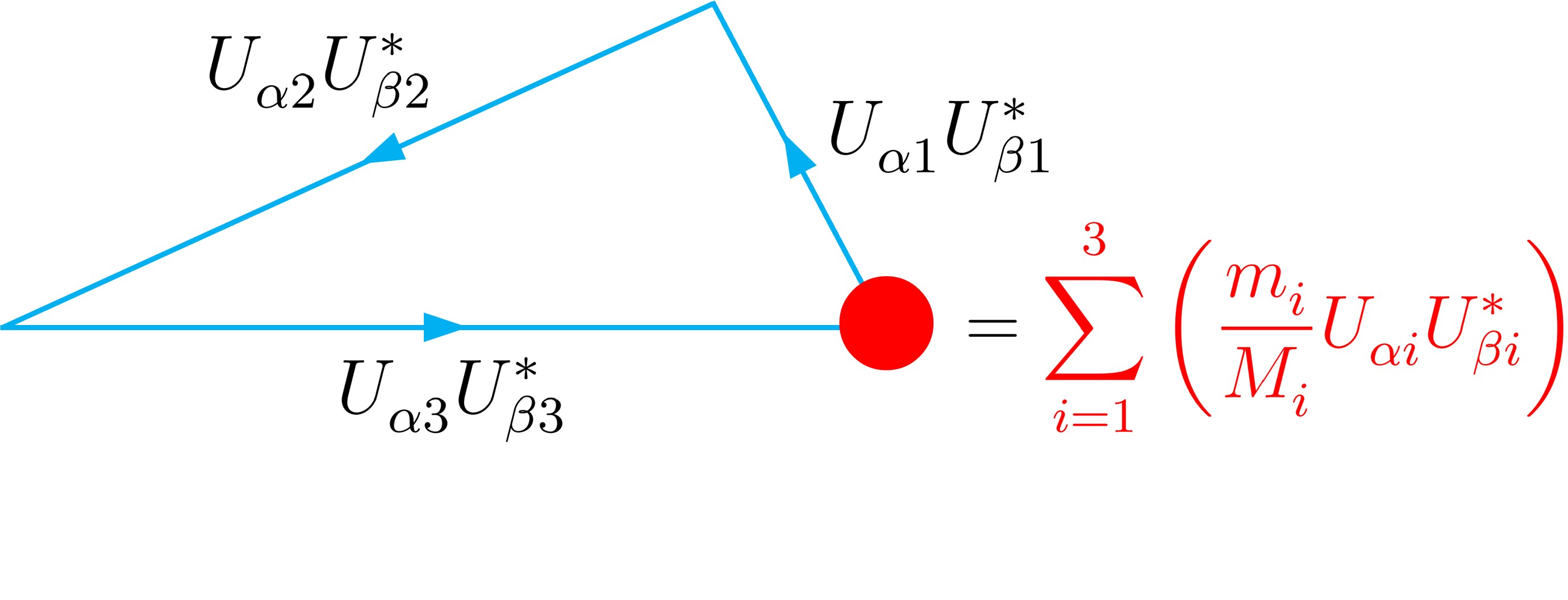}
\vspace{-1cm}
\caption{A schematic illustration of the FSS-associated unitarity
hexagons reduced to the {\it effective} triangles in the complex
plane (for $\alpha, \beta = e, \mu, \tau$ and $\alpha \neq \beta$).}
\label{fig2}
\end{center}
\end{figure}

Furthermore, the FSS relations help simplify the partial unitary condition
in Eq.~(\ref{5}) to the form
\begin{eqnarray}
\sum^3_{i=1} \left(1 + \frac{m^{}_i}{M^{}_i}\right)
U^{}_{\alpha i} U^{*}_{\beta i} = \delta^{}_{\alpha\beta} \; ,
\label{13}
%     (13)
\end{eqnarray}
implying that non-unitarity of the $3\times 3$ PMNS matrix $U$ should
be extremely small as a straightforward consequence of $m^{}_i/M^{}_i \ll 1$.
On the one hand, the unitarity hexagons fixed by the orthogonal conditions
from the off-diagonal parts of $U U^\dagger + R R^\dagger = {\bf 1}$
in the complex plane can all be reduced to the {\it effective}
unitarity triangles, as illustrated in Fig.~\ref{fig2}.
On the other hand, the normalization conditions
derived from the diagonal parts of $U U^\dagger + R R^\dagger = {\bf 1}$
can also be reduced to the {\it effective} normalization conditions
of $U$ itself to a good degree of accuracy.
%%%%%%%%%%%%%%%%%%%%%%%%%%%%%%% Figure 2 %%%%%%%%%%%%%%%%%%%%%%%%%%%%%%%
%%%%%%%%%%%%%%%%%%%%%%%%%%%%%%%%%%%%%%%%%%%%%%%%%%%%%%%%%%%%%%%%%%%%%%%%

\section{Parametrization of $U$ and $R$}

To be more explicit in describing the FSS scenario and exploring its
phenomenological implications, we take account of a complete
Euler-like parametrization of $U = A \cdot U^{}_0$ and 
$R$ as follows~\cite{Xing:2007zj,Xing:2011ur}:
\begin{eqnarray}
A \hspace{-0.2cm} & = & \hspace{-0.2cm}
\left( \begin{matrix} c^{}_{14} c^{}_{15} c^{}_{16} & 0 & 0
\cr \vspace{-0.47cm} \cr
\begin{array}{l} -c^{}_{14} c^{}_{15} \hat{s}^{}_{16} \hat{s}^*_{26} -
c^{}_{14} \hat{s}^{}_{15} \hat{s}^*_{25} c^{}_{26} \\
-\hat{s}^{}_{14} \hat{s}^*_{24} c^{}_{25} c^{}_{26} \end{array} &
c^{}_{24} c^{}_{25} c^{}_{26} & 0 \cr \vspace{-0.47cm} \cr
\begin{array}{l} -c^{}_{14} c^{}_{15} \hat{s}^{}_{16} c^{}_{26} \hat{s}^*_{36}
+ c^{}_{14} \hat{s}^{}_{15} \hat{s}^*_{25} \hat{s}^{}_{26} \hat{s}^*_{36} \\
- c^{}_{14} \hat{s}^{}_{15} c^{}_{25} \hat{s}^*_{35} c^{}_{36} +
\hat{s}^{}_{14} \hat{s}^*_{24} c^{}_{25} \hat{s}^{}_{26}
\hat{s}^*_{36} \\
+ \hat{s}^{}_{14} \hat{s}^*_{24} \hat{s}^{}_{25} \hat{s}^*_{35}
c^{}_{36} - \hat{s}^{}_{14} c^{}_{24} \hat{s}^*_{34} c^{}_{35}
c^{}_{36} \end{array} & \hspace{0.048cm}
\begin{array}{l} -c^{}_{24} c^{}_{25} \hat{s}^{}_{26} \hat{s}^*_{36} -
c^{}_{24} \hat{s}^{}_{25} \hat{s}^*_{35} c^{}_{36} \\
-\hat{s}^{}_{24} \hat{s}^*_{34} c^{}_{35} c^{}_{36} \end{array} \hspace{0.048cm}
& c^{}_{34} c^{}_{35} c^{}_{36} \cr \end{matrix} \right) \; ,
\nonumber \\
%---------------------------------------------------------------
R \hspace{-0.2cm} & = & \hspace{-0.2cm}
\left( \begin{matrix} \hat{s}^*_{14} c^{}_{15} c^{}_{16} &
\hat{s}^*_{15} c^{}_{16} & \hat{s}^*_{16} \cr \vspace{-0.47cm} \cr
\begin{array}{l} -\hat{s}^*_{14} c^{}_{15} \hat{s}^{}_{16} \hat{s}^*_{26} -
\hat{s}^*_{14} \hat{s}^{}_{15} \hat{s}^*_{25} c^{}_{26} \\
+ c^{}_{14} \hat{s}^*_{24} c^{}_{25} c^{}_{26} \end{array} & -
\hat{s}^*_{15} \hat{s}^{}_{16} \hat{s}^*_{26} + c^{}_{15}
\hat{s}^*_{25} c^{}_{26} & c^{}_{16} \hat{s}^*_{26} \cr \vspace{-0.47cm} \cr
\begin{array}{l} -\hat{s}^*_{14} c^{}_{15} \hat{s}^{}_{16} c^{}_{26}
\hat{s}^*_{36} + \hat{s}^*_{14} \hat{s}^{}_{15} \hat{s}^*_{25}
\hat{s}^{}_{26} \hat{s}^*_{36} \\ - \hat{s}^*_{14} \hat{s}^{}_{15}
c^{}_{25} \hat{s}^*_{35} c^{}_{36} - c^{}_{14} \hat{s}^*_{24}
c^{}_{25} \hat{s}^{}_{26}
\hat{s}^*_{36} \\
- c^{}_{14} \hat{s}^*_{24} \hat{s}^{}_{25} \hat{s}^*_{35}
c^{}_{36} + c^{}_{14} c^{}_{24} \hat{s}^*_{34} c^{}_{35} c^{}_{36}
\end{array} &
\begin{array}{l} -\hat{s}^*_{15} \hat{s}^{}_{16} c^{}_{26} \hat{s}^*_{36}
- c^{}_{15} \hat{s}^*_{25} \hat{s}^{}_{26} \hat{s}^*_{36} \\
+c^{}_{15} c^{}_{25} \hat{s}^*_{35} c^{}_{36} \end{array} &
c^{}_{16} c^{}_{26} \hat{s}^*_{36} \cr \end{matrix} \right) \; ,
\hspace{0.3cm}
\label{14}
%     (14)
\end{eqnarray}
where $c^{}_{ij} \equiv \cos\theta^{}_{ij}$, $s^{}_{ij} \equiv
\sin\theta^{}_{ij}$ and $\hat{s}^{}_{ij} \equiv s^{}_{ij}
e^{{\rm i}\delta^{}_{ij}}$ with $\theta^{}_{ij}$ and $\delta^{}_{ij}$
being the active-sterile flavor mixing angles and CP-violating phases
(for $i = 1, 2, 3$ and $j = 4, 5, 6$); and
\begin{eqnarray}
U^{}_0 \hspace{-0.2cm} & = & \hspace{-0.25cm}
\left( \begin{matrix} c^{}_{12} c^{}_{13} & \hat{s}^*_{12}
c^{}_{13} & \hat{s}^*_{13} \cr
-\hat{s}^{}_{12} c^{}_{23} -
c^{}_{12} \hat{s}^{}_{13} \hat{s}^*_{23} & c^{}_{12} c^{}_{23} -
\hat{s}^*_{12} \hat{s}^{}_{13} \hat{s}^*_{23} & c^{}_{13}
\hat{s}^*_{23} \cr
\hat{s}^{}_{12} \hat{s}^{}_{23} - c^{}_{12}
\hat{s}^{}_{13} c^{}_{23} & -c^{}_{12} \hat{s}^{}_{23} -
\hat{s}^*_{12} \hat{s}^{}_{13} c^{}_{23} & c^{}_{13} c^{}_{23}
\cr \end{matrix} \right)
\nonumber \\
\hspace{-0.2cm} & = & \hspace{-0.25cm}
\left( \begin{matrix}
e^{-{\rm i}\rho} & 0 & 0 \cr
0 & e^{-{\rm i}\sigma} & 0 \cr
0 & 0 & 1 \cr\end{matrix} \right) \hspace{-0.1cm}
\left( \begin{matrix} c^{}_{12} c^{}_{13} & s^{}_{12}
c^{}_{13} & s^{}_{13} e^{-{\rm i}\delta^{}_\nu} \cr
-s^{}_{12} c^{}_{23} - c^{}_{12} s^{}_{13} s^{}_{23} e^{{\rm i}\delta^{}_\nu}
& c^{}_{12} c^{}_{23} -
s^{}_{12} s^{}_{13} s^{}_{23} e^{{\rm i}\delta^{}_\nu} & c^{}_{13} s^{}_{23} \cr
s^{}_{12} s^{}_{23} - c^{}_{12} s^{}_{13} c^{}_{23} e^{{\rm i}\delta^{}_\nu}
& -c^{}_{12} s^{}_{23} - s^{}_{12} s^{}_{13} c^{}_{23} e^{{\rm i}\delta^{}_\nu}
& c^{}_{13} c^{}_{23}
\cr \end{matrix} \right) \hspace{-0.1cm}
\left( \begin{matrix}
e^{{\rm i}\rho} & 0 & 0 \cr
0 & e^{{\rm i}\sigma} & 0 \cr
0 & 0 & 1 \cr\end{matrix} \right) \; \hspace{0.6cm}
\label{15}
%     (15)
\end{eqnarray}
is the unitary part of the PMNS matrix $U$ for the three active Majorana neutrinos
which possess the CP-violating phases $\delta^{}_\nu \equiv \delta^{}_{13}
- \delta^{}_{12} - \delta^{}_{23}$, $\rho \equiv \delta^{}_{12} + \delta^{}_{23}$
and $\sigma \equiv \delta^{}_{23}$ in the standard phase convention as advocated
by the Particle Data Group~\cite{ParticleDataGroup:2024cfk}. One should keep in
mind that all the six flavor mixing parameters of $U^{}_0$ are {\it derivational}
in the sense that they can in principle be derived from the {\it original} seesaw
flavor parameters hidden in $D^{}_N$ and $R$~\cite{Xing:2024gmy}.

Given that the elements in the first rows of $U = A \cdot U^{}_0$ and
$R$ have much simpler expressions in this parametrization,
we may confront them with the FSS relations in Eq.~(\ref{7}) as an
easy example for the sake of illustration. We immediately arrive at
\begin{eqnarray}
\frac{m^{}_1}{M^{}_1} = \frac{t^2_{14}}{c^2_{12} c^2_{13}} \; ,
\quad
\frac{m^{}_2}{M^{}_2} = \frac{t^2_{15}}{s^2_{12} c^2_{13} c^2_{14}} \; ,
\quad
\frac{m^{}_3}{M^{}_3} = \frac{t^2_{16}}{s^2_{13} c^2_{14} c^2_{15}} \; ,
\hspace{0.3cm}
\label{16}
%     (16)
\end{eqnarray}
where $t^{}_{ij} \equiv \tan\theta^{}_{ij}$ is defined (for $i = 1, 2, 3$
and $j = 4, 5, 6$), and the phase conditions
$\delta^{}_{14} = \left(n + 1/2\right) \pi$,
$\delta^{}_{15} = \delta^{}_{12} + \left(n +1/2\right) \pi$ and
$\delta^{}_{16} = \delta^{}_{13} + \left(n + 1/2\right) \pi$ must be
satisfied (with $n$ being an arbitrary integer). These concise relations
allow us to determine or constrain $M^{}_i$ from $m^{}_i$, or vice versa,
once the relevant flavor mixing angles in the first rows of $U^{}_0$ and
$R$ are known or well constrained. In particular, Eq.~(\ref{16}) makes it
straightforward for us to obtain the phase-independent results
\begin{eqnarray}
&& \Sigma_\nu \equiv \sum^3_{i=1} m^{}_i =
\frac{t^2_{14}}{c^2_{12} c^2_{13}} M^{}_1 +
\frac{t^2_{15}}{s^2_{12} c^2_{13} c^2_{14}} M^{}_2 +
\frac{t^2_{16}}{s^2_{13} c^2_{14} c^2_{15}} M^{}_3 \; ,
\nonumber \\
&& \langle m\rangle^{}_\beta \equiv
\left[\sum^3_{i=1} m^2_i \left|U^{}_{e i}\right|^2\right]^{1/2}
= c^{}_{14} c^{}_{15} c^{}_{16}
\left[\frac{t^4_{14}}{c^2_{12} c^2_{13}} M^{2}_1 +
\frac{t^4_{15}}{s^2_{12} c^2_{13} c^4_{14}} M^{2}_2 +
\frac{t^4_{16}}{s^2_{13} c^4_{14} c^4_{15}} M^{2}_3\right]^{1/2} \; ,
\hspace{1.4cm}
\label{17}
%     (17)
\end{eqnarray}
which have respectively been constrained by the currently available data
accumulated from various cosmological observations and $\beta$-decay
experiments~\cite{ParticleDataGroup:2024cfk}; together with
\begin{eqnarray}
&& \Delta m^2_{21} \equiv m^2_2 - m^2_1 =
\frac{t^4_{15}}{s^4_{12} c^4_{13} c^4_{14}} M^2_2 -
\frac{t^4_{14}}{c^4_{12} c^4_{13}} M^2_1 \; ,
\nonumber \\
&& \Delta m^2_{31} \equiv m^2_3 - m^2_1 =
\frac{t^4_{16}}{s^4_{13} c^4_{14} c^4_{15}} M^2_3 -
\frac{t^4_{14}}{c^4_{12} c^4_{13}} M^2_1 \; ,
\hspace{1.5cm}
\label{18}
%     (18)
\end{eqnarray}
which have been determined to a remarkably good degree of accuracy (except
the sign of $\Delta m^2_{31}$)~\cite{ParticleDataGroup:2024cfk}. The sign
of $\Delta m^2_{31}$ is expected to be soon fixed in the undergoing JUNO
reactor antineutrino oscillation experiment. Then it will be possible to
obtain some much better constraints on the parameter space of $M^{}_i$
and $\theta^{}_{1j}$ (for $i = 1, 2, 3$ and $j = 4, 5, 6$) in the FSS
scenario. In comparison, the effective mass of the neutrinoless
double-beta ($0\nu2\beta$) decays involves two CP-violating phases
as follows
%%%%%%%%%%%%%%%%%%%%%%%%%%%%%%%%%%%%%%%%%%%%%%%%%%%%%%%%%%%%%%%%%%%%%%%%%
\footnote{In a general canonical seesaw framework with $M^{}_i \gg 
\langle \phi^0\rangle$ (for $i = 1, 2, 3$), the contributions of heavy Majorana 
neutrinos to $\langle m\rangle^{0\nu}_{2\beta}$ are highly suppressed and thus 
can be neglected~\cite{Xing:2009ce,Rodejohann:2009ve,Fang:2024hzy}. This case
is also valid in the FSS scenario.}:
%%%%%%%%%%%%%%%%%%%%%%%%%%%%%%%%%%%%%%%%%%%%%%%%%%%%%%%%%%%%%%%%%%%%%%%%%
\begin{eqnarray}
\langle m\rangle^{0\nu}_{2\beta} \equiv \left|\sum^3_{i=1} m^{}_i U^2_{e i}
\right| = \left|\sum^3_{i=1} M^{}_i R^2_{e i}
\right| = \left|M^{}_1 s^2_{14} c^2_{15} c^2_{16} + M^{}_2 s^2_{15} c^2_{16}
e^{{\rm i} 2\alpha^{}_1} + M^{}_3 s^2_{16} e^{{\rm i} 2\left(\alpha^{}_1
+ \beta^{}_1\right)} \right| \;
\label{19}
%     (19)
\end{eqnarray}
with $\alpha^{}_1 \equiv \delta^{}_{14} - \delta^{}_{15}$ and
$\beta^{}_1 \equiv \delta^{}_{15} - \delta^{}_{16}$.
The present upper bound on
$\langle m\rangle^{0\nu}_{2\beta}$~\cite{ParticleDataGroup:2024cfk} and
a future definite measurement of $\langle m\rangle^{0\nu}_{2\beta}$ may
help a lot to constrain the relevant heavy Majorana neutrino 
masses and active-sterile flavor mixing parameters.

It is worth pointing out that a combination of Eqs.~(\ref{17}) and
(\ref{18}) allows us to determine the masses of three heavy Majorana
neutrinos as follows:
\begin{eqnarray}
M^2_1 \hspace{-0.2cm} & = & \hspace{-0.2cm}
\frac{c^4_{12} c^4_{13}}{\left(s^2_{12} c^2_{13} + s^2_{13}\right)
t^4_{14}} \left[\frac{\langle m\rangle^2_\beta}{c^2_{14} c^2_{15}
c^2_{16}} - s^2_{12} c^2_{13} \Delta m^2_{21} - s^2_{13}
\Delta m^2_{31}\right] \; , \hspace{0.4cm}
\nonumber \\
M^2_2 \hspace{-0.2cm} & = & \hspace{-0.2cm}
\frac{s^4_{12} c^4_{13} c^4_{14}}{\left(s^2_{12} c^2_{13} + s^2_{13}\right)
t^4_{15}} \left[\frac{\langle m\rangle^2_\beta}{c^2_{14} c^2_{15}
c^2_{16}} - s^2_{13} \Delta m^2_{32}\right] \; ,
\nonumber \\
M^2_3 \hspace{-0.2cm} & = & \hspace{-0.2cm}
\frac{s^4_{13} c^4_{14} c^4_{15}}{\left(s^2_{12} c^2_{13} + s^2_{13}\right)
t^4_{16}} \left[\frac{\langle m\rangle^2_\beta}{c^2_{14} c^2_{15}
c^2_{16}} + s^2_{12} c^2_{13} \Delta m^2_{32}\right] \; .
\label{20}
%     (20)
\end{eqnarray}
The positivity of $M^2_i$ (for $i = 1, 2, 3$) requires that the term
$\langle m\rangle^2_\beta/\left(c^2_{14} c^2_{15} c^2_{16}\right)$ should
be larger than the other relevant terms in the square brackets on the
right-hand sides of the above three equations, no matter whether the
signs of $\Delta m^2_{31}$ and $\Delta m^2_{32}$ are positive (normal
neutrino mass ordering) or negative (inverted neutrino mass ordering).
This observation implies that the mass ordering of three heavy neutrinos
are closely correlated with that of three light neutrinos in the FSS
scenario, and thus it is possible to probe the former from the latter
if $\langle m\rangle^{}_\beta$ and $\theta^{}_{1j}$ (for $j = 4, 5, 6$)
are measured or sufficiently constrained in the future neutrino experiments.

Regarding the active-sterile flavor mixing angles, the limits
$(\theta^{}_{1j}, \theta^{}_{2j}, \theta^{}_{3j}) <
(2.92^\circ, 0.27^\circ, 2.56^\circ)$ in the $m^{}_1 < m^{}_2 < m^{}_3$
case or the bounds $(\theta^{}_{1j}, \theta^{}_{2j}, \theta^{}_{3j}) <
(3.03^\circ, 0.26^\circ, 2.31^\circ)$ in the $m^{}_3 < m^{}_1 < m^{}_2$
case have recently been derived for arbitrary values of $\delta^{}_{ij}$
(for $i = 1, 2, 3$ and $j = 4, 5, 6$)~\cite{Xing:2024gmy}, with the help
of rather stringent numerical constraints on $U U^\dagger = A A^\dagger$
obtained from a careful global analysis of the currently available 
electroweak precision measurements and neutrino oscillation data in the 
canonical seesaw framework~\cite{Blennow:2023mqx}. As a result,
\begin{eqnarray}
A \hspace{-0.2cm} & = & \hspace{-0.25cm}
\left(\begin{matrix} 1 - \frac{1}{2} \left(s^2_{14} +
s^2_{15} + s^2_{16}\right) & 0 & 0 \cr
- \left(\hat{s}^{}_{14} \hat{s}^*_{24} + \hat{s}^{}_{15} \hat{s}^*_{25}
+ \hat{s}^{}_{16} \hat{s}^*_{26}\right) &
1 - \frac{1}{2} \left(s^2_{24} + s^2_{25} + s^2_{26}\right)
& 0 \cr
- \left(\hat{s}^{}_{14} \hat{s}^*_{34} + \hat{s}^{}_{15}
\hat{s}^*_{35} + \hat{s}^{}_{16} \hat{s}^*_{36}\right) &
- \left(\hat{s}^{}_{24} \hat{s}^*_{34} + \hat{s}^{}_{25} \hat{s}^*_{35}
+ \hat{s}^{}_{26} \hat{s}^*_{36}\right) &
1- \frac{1}{2} \left(s^2_{34} + s^2_{35} + s^2_{36}\right) \cr
\end{matrix} \right) + {\cal O}\left(s^4_{ij}\right) \; ,
\nonumber \\
R \hspace{-0.2cm} & = & \hspace{-0.25cm}
\left( \begin{matrix} \hat{s}^*_{14} &
\hat{s}^*_{15} & \hat{s}^*_{16} \cr
\hat{s}^*_{24} & \hat{s}^*_{25} &
\hat{s}^*_{26} \cr
\hat{s}^*_{34} & \hat{s}^*_{35} &
\hat{s}^*_{36} \cr \end{matrix} \right) + {\cal O}\left(s^3_{ij}\right) \; ,
\label{21}
%     (21)
\end{eqnarray}
where ${\cal O}\left(s^3_{ij}\right) < 1.4 \times 10^{-4}$ and
${\cal O}\left(s^4_{ij}\right) < 7.8 \times 10^{-6}$ (for $i = 1, 2, 3$
and $j = 4, 5, 6$). So non-unitarity of the PMNS matrix $U$ as
characterized by the deviation of $A$ from $\bf 1$ is at most of
${\cal O}\left(s^2_{ij}\right) < 2.8 \times 10^{-3}$~\cite{Xing:2025bdm},
and the leading terms of $A$ and $R$ in Eq.~(\ref{21}) should be very
good approximations for the study of phenomenology of
the seesaw mechanism, including the FSS scenario.

\section{Analytical approximations}

Given the very fact that the experiments of normal neutrino oscillations
have been playing the most important role in determining the
fundamental flavor parameters of three active Majorana neutrinos, we are
going to calculate the three flavor mixing angles and the three
CP-violating phases of $U^{}_0$ in the leading-order approximations of
$R$ and $U$. First of all, we obtain the following results by considering
the first rows and third columns of $U = A \cdot U^{}_0$ and $R$:
\begin{eqnarray}
s^{}_{13} \hspace{-0.2cm} & \simeq & \hspace{-0.2cm}
\left|U^{}_{e 3}\right| \simeq \sqrt{\frac{M^{}_3}{m^{}_3}}
\hspace{0.07cm} s^{}_{16} \; ,
\nonumber \\
t^{}_{12} \hspace{-0.2cm} & \simeq & \hspace{-0.2cm}
\left|\frac{U^{}_{e 2}}{U^{}_{e 1}}\right| \simeq
\sqrt{\frac{m^{}_1 M^{}_2}{m^{}_2 M^{}_1}} \cdot
\frac{s^{}_{15}}{s^{}_{14}} \; , \hspace{0.4cm}
\nonumber \\
t^{}_{23} \hspace{-0.2cm} & \simeq & \hspace{-0.2cm}
\left|\frac{U^{}_{\mu 3}}{U^{}_{\tau 3}}\right| \simeq
\frac{s^{}_{26}}{s^{}_{36}} \; .
\label{22}
%     (22)
\end{eqnarray}
One can see that the expression of $t^{}_{23}$ is independent of the ratios
of light and heavy neutrino masses, and thus $\theta^{}_{23} \simeq \pi/4$
in the $\mu$-$\tau$ reflection symmetry limit implies $\theta^{}_{26}
\simeq \theta^{}_{36}$ or equivalently $\left|R^{}_{\mu 3}\right| \simeq
\left|R^{}_{\tau 3}\right|$ thanks to the exact seesaw relationship between
$U$ and $R$~\cite{Xing:2022oob}.

To calculate the CP-violating phase $\delta^{}_\nu$ of $U^{}_0$ defined
in Eq.~(\ref{15}), we first calculate the rephasing invariants of
CP violation ${\cal J}^{ij}_{\alpha\beta}$ defined in Eq. (\ref{11})
in the leading-order approximations of $A$ and $R$:
\begin{eqnarray}
{\cal J}^{12}_{e\mu} \hspace{-0.2cm} & \simeq & \hspace{-0.2cm}
\frac{M^{}_1 M^{}_2}{m^{}_1 m^{}_2} s^{}_{14} s^{}_{15}
s^{}_{24} s^{}_{25} \sin\left(\alpha^{}_2 - \alpha^{}_1\right) \; ,
\nonumber \\
{\cal J}^{23}_{e\mu} \hspace{-0.2cm} & \simeq & \hspace{-0.2cm}
\frac{M^{}_2 M^{}_3}{m^{}_2 m^{}_3} s^{}_{15} s^{}_{16}
s^{}_{25} s^{}_{26} \sin\left(\beta^{}_2 - \beta^{}_1\right) \; ,
\nonumber \\
{\cal J}^{31}_{e\mu} \hspace{-0.2cm} & \simeq & \hspace{-0.2cm}
\frac{M^{}_1 M^{}_3}{m^{}_1 m^{}_3} s^{}_{14} s^{}_{16}
s^{}_{24} s^{}_{26} \sin\left(\gamma^{}_2 - \gamma^{}_1\right) \; ;
%--------------------------------------------------------------------
\nonumber \\
{\cal J}^{12}_{\mu\tau} \hspace{-0.2cm} & \simeq & \hspace{-0.2cm}
\frac{M^{}_1 M^{}_2}{m^{}_1 m^{}_2} s^{}_{24} s^{}_{25}
s^{}_{34} s^{}_{35} \sin\left(\alpha^{}_3 - \alpha^{}_2\right) \; ,
\nonumber \\
{\cal J}^{23}_{\mu\tau} \hspace{-0.2cm} & \simeq & \hspace{-0.2cm}
\frac{M^{}_2 M^{}_3}{m^{}_2 m^{}_3} s^{}_{25} s^{}_{26}
s^{}_{35} s^{}_{36} \sin\left(\beta^{}_3 - \beta^{}_2\right) \; ,
\nonumber \\
{\cal J}^{31}_{\mu\tau} \hspace{-0.2cm} & \simeq & \hspace{-0.2cm}
\frac{M^{}_1 M^{}_3}{m^{}_1 m^{}_3} s^{}_{24} s^{}_{26}
s^{}_{34} s^{}_{36} \sin\left(\gamma^{}_3 - \gamma^{}_2\right) \; ;
%--------------------------------------------------------------------
\nonumber \\
{\cal J}^{12}_{\tau e} \hspace{-0.2cm} & \simeq & \hspace{-0.2cm}
\frac{M^{}_1 M^{}_2}{m^{}_1 m^{}_2} s^{}_{14} s^{}_{15}
s^{}_{34} s^{}_{35} \sin\left(\alpha^{}_1 - \alpha^{}_3\right) \; ,
\nonumber \\
{\cal J}^{23}_{\tau e} \hspace{-0.2cm} & \simeq & \hspace{-0.2cm}
\frac{M^{}_2 M^{}_3}{m^{}_2 m^{}_3} s^{}_{15} s^{}_{16}
s^{}_{35} s^{}_{36} \sin\left(\beta^{}_1 - \beta^{}_3\right) \; ,
\nonumber \\
{\cal J}^{31}_{\tau e} \hspace{-0.2cm} & \simeq & \hspace{-0.2cm}
\frac{M^{}_1 M^{}_3}{m^{}_1 m^{}_3} s^{}_{14} s^{}_{16}
s^{}_{34} s^{}_{36} \sin\left(\gamma^{}_1 - \gamma^{}_3\right) \; ,
\hspace{0.6cm}
\label{23}
%     (23)
\end{eqnarray}
where the CP-violating phases
\begin{eqnarray}
\alpha^{}_i \equiv \delta^{}_{i 4} - \delta^{}_{i 5} \; , \quad
\beta^{}_i \equiv \delta^{}_{i5} - \delta^{}_{i6} \; , \quad
\gamma^{}_i \equiv \delta^{}_{i6} - \delta^{}_{i4} \;
\label{24}
%     (24)
\end{eqnarray}
satisfy the sum rule $\alpha^{}_i + \beta^{}_i + \gamma^{}_i = 0$ (for
$i = 1, 2, 3$)~\cite{Xing:2024xwb}. In this approximation
(i.e., $U \to U^{}_0$), the above nine rephasing invariants should be
equal to one another and all approach a universal Jarlskog invariant of 
leptonic CP violation of the form~\cite{Jarlskog:1985ht,Wu:1985ea}
\begin{eqnarray}
{\cal J}^{}_0 = c^{}_{12} s^{}_{12} c^2_{13} s^{}_{13} c^{}_{23} s^{}_{23}
\sin\delta^{}_\nu \; ,
\label{25}
%     (25)
\end{eqnarray}
where $\delta^{}_\nu$ has been defined in Eq.~(\ref{15}) and is responsible
for the dominant effects of CP violation in normal neutrino
oscillations~\cite{Barger:1980jm}. Then the equalities
${\cal J}^{}_1 \simeq {\cal J}^{}_4 \simeq {\cal J}^{}_7$,
${\cal J}^{}_2 \simeq {\cal J}^{}_5 \simeq {\cal J}^{}_8$ and
${\cal J}^{}_3 \simeq {\cal J}^{}_6 \simeq {\cal J}^{}_9$ allow us to
arrive at the following phase correlations:
\begin{eqnarray}
s^{}_{14} s^{}_{15} \sin\left(\alpha^{}_2 - \alpha^{}_1\right) \
\hspace{-0.2cm} & \simeq & \hspace{-0.2cm}
s^{}_{34} s^{}_{35} \sin\left(\alpha^{}_3 - \alpha^{}_2\right) \; ,
\nonumber \\
s^{}_{14} s^{}_{15} \sin\left(\alpha^{}_1 - \alpha^{}_3\right) \
\hspace{-0.2cm} & \simeq & \hspace{-0.2cm}
s^{}_{24} s^{}_{25} \sin\left(\alpha^{}_3 - \alpha^{}_2\right) \; ,
\nonumber \\
s^{}_{24} s^{}_{25} \sin\left(\alpha^{}_2 - \alpha^{}_1\right) \
\hspace{-0.2cm} & \simeq & \hspace{-0.2cm}
s^{}_{34} s^{}_{35} \sin\left(\alpha^{}_1 - \alpha^{}_3\right) \; ;
\nonumber \\
%--------------------------------------------------------------------
s^{}_{15} s^{}_{16} \sin\left(\beta^{}_2 - \beta^{}_1\right) \
\hspace{-0.2cm} & \simeq & \hspace{-0.2cm}
s^{}_{35} s^{}_{36} \sin\left(\beta^{}_3 - \beta^{}_2\right) \; ,
\nonumber \\
s^{}_{15} s^{}_{16} \sin\left(\beta^{}_1 - \beta^{}_3\right) \
\hspace{-0.2cm} & \simeq & \hspace{-0.2cm}
s^{}_{25} s^{}_{26} \sin\left(\beta^{}_3 - \beta^{}_2\right) \; ,
\nonumber \\
s^{}_{25} s^{}_{26} \sin\left(\beta^{}_2 - \beta^{}_1\right) \
\hspace{-0.2cm} & \simeq & \hspace{-0.2cm}
s^{}_{35} s^{}_{36} \sin\left(\beta^{}_1 - \beta^{}_3\right) \; ;
\nonumber \\
%--------------------------------------------------------------------
s^{}_{14} s^{}_{16} \sin\left(\gamma^{}_2 - \gamma^{}_1\right) \
\hspace{-0.2cm} & \simeq & \hspace{-0.2cm}
s^{}_{34} s^{}_{36} \sin\left(\gamma^{}_3 - \gamma^{}_2\right) \; ,
\nonumber \\
s^{}_{14} s^{}_{16} \sin\left(\gamma^{}_1 - \gamma^{}_3\right) \
\hspace{-0.2cm} & \simeq & \hspace{-0.2cm}
s^{}_{24} s^{}_{26} \sin\left(\gamma^{}_3 - \gamma^{}_2\right) \; ,
\nonumber \\
s^{}_{24} s^{}_{26} \sin\left(\gamma^{}_2 - \gamma^{}_1\right) \
\hspace{-0.2cm} & \simeq & \hspace{-0.2cm}
s^{}_{34} s^{}_{36} \sin\left(\gamma^{}_1 - \gamma^{}_3\right) \; ,
\hspace{0.75cm}
\label{26}
%     (26)
\end{eqnarray}
implying that there are only two independent phase differences (e.g.,
$\alpha^{}_2 - \alpha^{}_1$ and $\beta^{}_2 - \beta^{}_1$) in Eq.~(\ref{23}).
But if ${\cal J}^{12}_{e\mu} \simeq {\cal J}^{23}_{e\mu}$ is
also taken into account, one will be left with
\begin{eqnarray}
\sin\left(\beta^{}_2 - \beta^{}_1\right) \simeq
\frac{M^{}_1 m^{}_3 s^{}_{14} s^{}_{24}}{m^{}_1 M^{}_3
s^{}_{16} s^{}_{26}} \sin\left(\alpha^{}_2 - \alpha^{}_1\right) \; .
\label{27}
%     (27)
\end{eqnarray}
So the nine rephasing invariants of CP violation under discussion
depend only on a single independent phase difference
$\left(\alpha^{}_2 - \alpha^{}_1\right)$ in the FSS scenario.

Substituting Eq.~(\ref{22}) into Eq.~(\ref{25}) and taking
${\cal J}^{12}_{e\mu} \simeq {\cal J}^{}_0$ in the same approximation,
we obtain the result for the Dirac-like CP-violating phase
$\delta^{}_\nu$ as follows:
\begin{eqnarray}
\sin\delta^{}_\nu \simeq \left(\frac{m^{}_3}{M^{}_3} - s^2_{16}\right)
\sqrt{\frac{M^{}_1 M^{}_2 m^{}_3}{m^{}_1 m^{}_2 M^{}_3}} \cdot
\frac{s^{}_{24} s^{}_{25}}{s^{}_{16} s^{}_{26} s^{}_{36}}
\sin\left(\alpha^{}_2 - \alpha^{}_1\right) \; .
\label{28}
%     (28)
\end{eqnarray}
In comparison, the two Majorana-like CP-violating phases of $U^{}_0$,
namely $\rho$ and $\sigma$ as defined in Eq.~(\ref{15}), can be directly
extracted from the expression of $\langle m\rangle^{0\nu}_{2\beta}$ in
Eq.~(\ref{19}):
\begin{eqnarray}
\rho = -\left(\alpha^{}_1 + \beta^{}_1 + \delta^{}_\nu\right) \; ,
\quad \sigma = -\beta^{}_1 - \delta^{}_\nu \; ,
\label{29}
%     (29)
\end{eqnarray}
in which $\delta^{}_\nu$ is a function of $\alpha^{}_2 - \alpha^{}_1$ as
shown in Eq.~(\ref{28}). Eqs.~(\ref{22}), (\ref{28}) and (\ref{29}) tell
us that the flavor mixing parameters of $U^{}_0$ are easily calculable
in terms of the original seesaw flavor parameters in the FSS scenario,
making the latter more easily testable in the upcoming neutrino oscillation
experiments.

Note that the CP-violating asymmetry $\varepsilon^{}_i$ associated with the
lepton-number-violating decays of $N^{}_i$ depends on the rephasing
invariants ${\cal V}^{ij}_{\alpha\beta}$ (for $i, j = 1, 2, 3$ and
$\alpha, \beta = e, \mu, \tau$), as indicated in Eq.~(\ref{12}). In the
leading-order approximation of $R$, we obtain
\begin{eqnarray}
\left({\cal V}^{12}_{ee} \; , \; {\cal V}^{12}_{\mu\mu} \; , \;
{\cal V}^{12}_{\tau\tau}\right)  
\hspace{-0.2cm} & \simeq & \hspace{-0.2cm}
- \frac{M^{}_1 M^{}_2}{m^{}_1 m^{}_2} \left(s^2_{14} s^2_{15}
\sin 2\alpha^{}_1 \; , \; s^2_{24} s^2_{25} \sin 2\alpha^{}_2 \; ,
\; s^2_{34} s^2_{35} \sin 2\alpha^{}_3\right) \; , \hspace{0.4cm}
%--------------------------------------------------------------------
\nonumber \\
\left({\cal V}^{23}_{ee} \; , \; {\cal V}^{23}_{\mu\mu} \; , \;
{\cal V}^{23}_{\tau\tau}\right)
\hspace{-0.2cm} & \simeq & \hspace{-0.2cm}
- \frac{M^{}_2 M^{}_3}{m^{}_2 m^{}_3} \left(s^2_{15} s^2_{16}
\sin 2\beta^{}_1 \; , \; s^2_{25} s^2_{26} \sin 2\beta^{}_2 \; ,
\; s^2_{35} s^2_{36} \sin 2\beta^{}_3\right) \; , \hspace{0.4cm}
%--------------------------------------------------------------------
\nonumber \\
\left({\cal V}^{31}_{ee} \; , \; {\cal V}^{31}_{\mu\mu} \; ,
\; {\cal V}^{31}_{\tau\tau}\right)
\hspace{-0.2cm} & \simeq & \hspace{-0.2cm}
- \frac{M^{}_1 M^{}_3}{m^{}_1 m^{}_3} \left(s^2_{14} s^2_{16}
\sin 2\gamma^{}_1 \; , \; s^2_{24} s^2_{26} \sin 2\gamma^{}_2 \; ,
\; s^2_{34} s^2_{36} \sin 2\gamma^{}_3\right) \; ; \hspace{0.4cm}
\label{30}
%     (30)
\end{eqnarray}
together with 
\begin{eqnarray}
{\cal V}^{12}_{e\mu} \hspace{-0.2cm} & \simeq & \hspace{-0.2cm}
-\frac{M^{}_1 M^{}_2}{m^{}_1 m^{}_2} s^{}_{14} s^{}_{15}
s^{}_{24} s^{}_{25} \sin\left(\alpha^{}_1 + \alpha^{}_2\right) \; ,
\nonumber \\
{\cal V}^{23}_{e\mu} \hspace{-0.2cm} & \simeq & \hspace{-0.2cm}
-\frac{M^{}_2 M^{}_3}{m^{}_2 m^{}_3} s^{}_{15} s^{}_{16}
s^{}_{25} s^{}_{26} \sin\left(\beta^{}_1 + \beta^{}_2\right) \; ,
\nonumber \\
{\cal V}^{31}_{e\mu} \hspace{-0.2cm} & \simeq & \hspace{-0.2cm}
-\frac{M^{}_1 M^{}_3}{m^{}_1 m^{}_3} s^{}_{14} s^{}_{16}
s^{}_{24} s^{}_{26} \sin\left(\gamma^{}_1 + \gamma^{}_2\right) \; ;
%--------------------------------------------------------------------
\nonumber \\
{\cal V}^{12}_{\mu\tau} \hspace{-0.2cm} & \simeq & \hspace{-0.2cm}
-\frac{M^{}_1 M^{}_2}{m^{}_1 m^{}_2} s^{}_{24} s^{}_{25}
s^{}_{34} s^{}_{35} \sin\left(\alpha^{}_2 + \alpha^{}_3\right) \; ,
\nonumber \\
{\cal V}^{23}_{\mu\tau} \hspace{-0.2cm} & \simeq & \hspace{-0.2cm}
-\frac{M^{}_2 M^{}_3}{m^{}_2 m^{}_3} s^{}_{25} s^{}_{26}
s^{}_{35} s^{}_{36} \sin\left(\beta^{}_2 + \beta^{}_3\right) \; ,
\nonumber \\
{\cal V}^{31}_{\mu\tau} \hspace{-0.2cm} & \simeq & \hspace{-0.2cm}
-\frac{M^{}_1 M^{}_3}{m^{}_1 m^{}_3} s^{}_{24} s^{}_{26}
s^{}_{34} s^{}_{36} \sin\left(\gamma^{}_2 + \gamma^{}_3\right) \; ;
%--------------------------------------------------------------------
\nonumber \\
{\cal V}^{12}_{\tau e} \hspace{-0.2cm} & \simeq & \hspace{-0.2cm}
-\frac{M^{}_1 M^{}_2}{m^{}_1 m^{}_2} s^{}_{14} s^{}_{15}
s^{}_{34} s^{}_{35} \sin\left(\alpha^{}_1 + \alpha^{}_3\right) \; ,
\nonumber \\
{\cal V}^{23}_{\tau e} \hspace{-0.2cm} & \simeq & \hspace{-0.2cm}
-\frac{M^{}_2 M^{}_3}{m^{}_2 m^{}_3} s^{}_{15} s^{}_{16}
s^{}_{35} s^{}_{36} \sin\left(\beta^{}_1 + \beta^{}_3\right) \; ,
\nonumber \\
{\cal V}^{31}_{\tau e} \hspace{-0.2cm} & \simeq & \hspace{-0.2cm}
-\frac{M^{}_1 M^{}_3}{m^{}_1 m^{}_3} s^{}_{14} s^{}_{16}
s^{}_{34} s^{}_{36} \sin\left(\gamma^{}_1 + \gamma^{}_3\right) \; ,
\hspace{0.7cm}
\label{31}
%     (31)
\end{eqnarray}
with the help of Eq.~(\ref{11}). It becomes obvious that the phase 
combinations of these eighteen invariants are different from those of
${\cal J}^{ij}_{\alpha\beta}$, because they are respectively associated
with CP violation in the lepton-number-violating processes and that in
the lepton-number-conserving processes.  

\section{Summary}

Different from the traditional model building exercises based on the
canonical seesaw mechanism, which have been done in a chosen flavor basis of
the Majorana neutrinos, the FSS scenario proposed in the present
work originates from a particular but brand new solution to the exact seesaw
equation in the mass basis of the neutrinos. The most salient feature
of this novel idea is that each family of the light and heavy Majorana
neutrinos has its own seesaw formula of the form $m^{}_i/M^{}_i =
-R^{2}_{\alpha i}/U^{2}_{\alpha i}$ (for $i = 1, 2, 3$ and $\alpha =
e, \mu, \tau$), making it possible to reduce the nonlinear entanglement 
of the seesaw parameters and thus enhance the calculability of the seesaw 
mechanism. We have concretely shown that the FSS scenario allows us to 
establish simple and testable relations between the original seesaw flavor 
parameters and those light and active degrees of freedom.

Given the fact that the canonical seesaw mechanism is the most
natural and popular theoretical framework for the origin of tiny neutrino
masses beyond the SM but its predictability and testability are rather
limited due to the unknown flavor structures, we stress that the FSS
scenario as an especially  novel benchmark example deserves a thorough
study in the era of precision neutrino physics.

\vspace{0.3cm}

{\it The author is indebted to the CERN Department of Theoretical Physics,
where the FSS idea came to him during a short-term visit in May 2026. He
is also grateful to Y.F. Li, D. Zhang, Z.H. Zhao and S. Zhou for their
helpful comments. This work was supported in part by the National Natural 
Science Foundation of China under Grant No. 12535007 and by the Scientific and
Technological Innovation Program of the Institute of High Energy Physics
under Grant No. E55457U2}.

\end{document}